\documentclass[10pt,conference]{IEEEtran}
\usepackage[english]{babel} 
\newcommand{\B}[1]{\boldsymbol{#1}}

\usepackage{amsmath}
\usepackage{amssymb}
\usepackage{epsfig}
\usepackage{psfrag}

\begin{document}
\selectlanguage{english}
\title{{Analysis of 1-bit Output Noncoherent Fading Channels in the Low SNR Regime}\vspace{-0.2cm}}
\author{{Amine Mezghani and Josef A. Nossek}
\authorblockA{\\Institute for Circuit Theory and Signal Processing\\ Munich University of
Technology, 80290 Munich, Germany\\
E-Mail: \{Mezghani, Nossek\}@nws.ei.tum.de}\vspace{-0.8cm}}
\maketitle 
\begin{abstract}
We consider  general multi-antenna fading channels with coarsely quantized outputs, where the channel is unknown  to the transmitter and receiver. This analysis is of interest in the context of sensor network communication where  low power and low cost are key requirements (e.g. standard IEEE 802.15.4 applications). This is also motivated by highly energy constrained communications devices where
sampling the signal may be more energy consuming than processing or transmitting it. Therefore the analog-to-digital converters (ADCs) for such applications should be low-resolution, in order to reduce their cost and power consumption. In this paper, we consider the extreme case of only 1-bit ADC for each receive signal component. We derive asymptotics of  the mutual information up to the second order in the signal-to-noise ratio (SNR) under average and peak power constraints   and study the impact of quantization. We show that up to second order in SNR, the mutual information of a   system with two-level (sign) output signals incorporates only a power penalty factor of almost $\frac{\pi}{2}$ (1.96 dB) compared to the system with infinite resolution for all channels of practical interest. This generalizes a recent result for the coherent case \cite{mezghaniisit2007}. 
\end{abstract}
\section{Introduction}
Several contributions studied MIMO channels operating in Rayleigh fading environments,  
 especially in the low SNR \cite{verdu,prelov,rao,wu,sethuraman,durisi} and high SNR \cite{zheng2} regime. Unfortunately, most of these contributions assume that the receiver has access to the channel data with infinite precision. In practice, however, a quantizer (analog-to-digital converter) is applied to the receive signal, so that the channel measurements can be processed in the digital domain. 

The relative simplicity of realizing and integrating functions digitally make it desirable, to move the analog-to-digital interface further towards the antenna.
 However, both the resolution and the speed required of the
ADC tend to rise, making them expensive, power intensive and even infeasible, especially in ultra-wideband and/or high speed applications.
 In fact, in order to reduce circuit complexity and save power and area, low resolution ADCs have to be employed \cite{wentzloff}. 
 
In \cite{mezghaniisit2007,nossek}, we study the effects of quantization from an information theoretical point of  view for  MIMO systems, where the channel is perfectly known at the receiver. It turns out that the loss in channel capacity due
to coarse quantization is surprisingly small at low to moderate SNR. In \cite{mezghaniisit2008} the block fading SISO noncoherent channel was studied in details. 
 Motivated by these works, we aim to study the communication performance of general noncoherent fading channels taking into account the coarse quantization. We consider the extreme case of 1-bit quantized (hard-decision)  MIMO channel with  no CSI at the transmitter and the receiver. 
When a single bit hard-limiter is used, the implementation of the
all digital receiver is considerably simplified \cite{hoyos,donnell,blazquez}. In particular,  automatic gain control (AGC), linearity requirements of RF components and  multipliers for signal correlation are no more necessary.

\par Our paper is organized as follows. Section \ref{section:scmodel} describes the system model and notational issues. In Section \ref{section:mutual} we give the general expression of the mutual information between the inputs and the quantized outputs of the MIMO system, then we expand it into a Taylor series up to the second order of the SNR  in Section \ref{section:mutual2}. Finally, in Section \ref{section:receiver}, we utilize these results to elaborate on the asymptotic capacity of 1-bit multiple receive antennas (SIMO) channels with delay spread and receive correlation in a Rayleigh flat-fading environment.
\label{section:introduction}
\section{System Model And Notation}
\label{section:scmodel}
\begin{figure}[h]
\begin{center}
\psfrag{H}[c][c]{$\boldsymbol{H}$}
\psfrag{G}[c][c]{$\boldsymbol{G}$}
\psfrag{xd}[c][c]{$\boldsymbol{\hat{x}}$}
\psfrag{x}[c][c]{$\boldsymbol{x}$}
\psfrag{y}[c][c]{$\boldsymbol{y}$}
\psfrag{r}[c][c]{$\boldsymbol{r}$}
\psfrag{Q[.]}[c][c]{$Q(\bullet)$}
\psfrag{n}[c][c]{$\boldsymbol{\eta}$}
\psfrag{M}[c][c]{$M$}
\psfrag{N}[c][c]{$N$}
\psfrag{SNR}[c][c]{$\sqrt{\rho}$}
{\epsfig {file=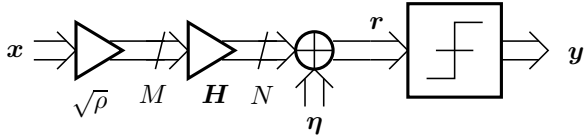, width = 7.5cm}}
\caption{One-bit Quantized MIMO System}
\label{downlink_figure}
\end{center}
\end{figure}

We consider a  point-to-point quantized MIMO channel with $M$ transmit dimensions (e.g. antennas) and $N$ receive dimensions at the  receiver. Fig.~\ref{downlink_figure} shows the general form of a quantized MIMO system, where $\boldsymbol{H} \in \mathbb{C}^{N\times M}$ is a random channel matrix. The entries of this channel matrix  are jointly complex circular Gaussian with zero-mean and possible mutual correlations. The channel realizations are unknown to both the transmitter and receiver. The vector $\boldsymbol{x} \in \mathbb{C}^{M}$ comprises the $M$ transmitted symbols, assumed to be subjected to a some average power constraint and peak power constraint. The vector $\boldsymbol{\eta}$ represents the additive noise, whose entries are i.i.d. and distributed as $\mathcal{CN}(0,1)$.  $\boldsymbol{r}\in\mathbb{C}^{N}$ is the unquantized channel output
\begin{equation}
 \boldsymbol{r}=\sqrt{\rho}\boldsymbol{H}\boldsymbol{x}+\boldsymbol{\eta},
\end{equation}
where $\rho$ represents the signal-to-noise ratio.  In our system, the real parts $r_{i,R}$ and the imaginary parts $r_{i,I}$ of the receive signals $r_i$, $1\leq i\leq N$, are each quantized by a $1$-bit resolution quantizer. Thus, the resulting
   quantized signals read as 
\begin{equation}
y_{i,c}=\textrm{sign}(r_{i,c})\in\{-1,1\},\textrm{ for }c\in\{R,I\},\textrm{ }1\leq i\leq N.
\end{equation}

\vspace{-0.2cm}
\vspace{-0.2cm}
Throughout our paper, $a_i$ denotes the $i$-th element of the vector $\boldsymbol{a}$ and $[\boldsymbol{a}]_{i,c}=a_{i,c}$ with $c\in\{R,I\}$ is the real or imaginary part of $a_i$. The operators $(\cdot)^\textrm{H}$ and $\textrm{tr}(\cdot)$ stand for Hermitian transpose and trace of a matrix, respectively. Vectors and matrices are denoted by lower and upper case italic bold letters, respectively.


\section{Mutual Information}
\label{section:mutual}
The mutual information (in nats/s/Hz) between the channel input and the quantized output in Fig.~\ref{downlink_figure} reads as \cite{cover}
\begin{equation}
\begin{aligned}	
I(\boldsymbol{x},\boldsymbol{y})&=H(\B{y})-H(\B{y}|\B{x})\\
&=\textrm{E}_{\boldsymbol{x}}\left[\sum_{\boldsymbol{y}}P(\boldsymbol{y}|\boldsymbol{x})\textrm{ln}\frac{P(\boldsymbol{y}|\boldsymbol{x})}{P(\boldsymbol{y})}\right],
\label{transinfo}
\end{aligned}
\end{equation}
with $P(\boldsymbol{y})=\textrm{E}_{\boldsymbol{x}}[P(\boldsymbol{y}|\boldsymbol{x})]$ and $\textrm{E}_{\boldsymbol{x}}[\cdot]$ is the expectation taken with respect to $\boldsymbol{x}$. Herewith, $H(\cdot)$ and $H(\cdot|\cdot)$ represent the entropy and the conditional entropy, respectively. 
Given the input $\B{x}$, the unquantized output $\B{r}$ is zero-mean complex Gaussian with covariance  $\textrm{E}[\B{r}\B{r}^{\rm H}|\B{x}]=(\textbf{I}_N+\rho{\rm E}_{\B{H}}[\B{H}\B{x}\B{x}^{\rm H} \B{H}^{\rm H}]) $, and thus we have   \\
\begin{equation}
   p(\B{r}|\B{x})=\frac{{\rm exp}(-\B{r}^{\rm H}(\textbf{I}_N+ \rho\rm{E}_{\B{H}}[\B{H}\B{x}\B{x}^{\rm H} \B{H}^{\rm H}])^{-1}\B{r})}{\pi^{N}{\rm det}(\textbf{I}_N+\rho\rm{E}_{\B{H}}[\B{H}\B{x}\B{x}^{\rm H} \B{H}^{\rm H}])}.
   \label{unq_cond_pro}
\end{equation}
Afterwards, we can express the conditional probability of the quantized output as
\begin{equation}
\begin{aligned}
&P(\boldsymbol{y}|\boldsymbol{x})=\int_0^{\infty}\!\!\!\!\!\!\!\cdots\!\!\!\int_0^{\infty} p(\boldsymbol{y} \bullet \B{r}|\B{x}) {\rm d}\B{r}=\\
&\int_0^{\infty}\!\!\!\!\!\!\!\!\!\cdots \!\!\! \int_0^{\infty}\!\!\!\!  \frac{{\rm exp}(-(\B{y} \bullet \boldsymbol{r} )^{\rm H}(\textbf{I}_N+ \rho\rm{E}[\B{H}\B{x}\B{x}^{\rm H} \B{H}^{\rm H}])^{-1}(\B{y} \bullet \boldsymbol{r}))}{\pi^{N}{\rm det}(\textbf{I}_N+\rho\rm{E}[\B{H}\B{x}\B{x}^{\rm H} \B{H}^{\rm H}])}{\rm d}\B{r},
\label{cond_pro}
\end{aligned}
\end{equation}
where the integration is performed over the positive orthant of the complex hyperplane and $\boldsymbol{y}\bullet \B{r}$ denotes an dimension-wise   vector product with $[\boldsymbol{y}\bullet \B{r}]_{i,c}=y_{i,c}r_{i,c}$. \\
The evaluation of this multiple integral is intractable in general. Thus let consider first a  simple lower bound involving the mutual information under perfect channel state information at the receiver studied in \cite{mezghaniisit2007}. By the chain rule and the non-negativity of the mutual information 
\begin{equation}
\begin{aligned}	
I(\boldsymbol{x},\boldsymbol{y})&=I(\boldsymbol{H},\boldsymbol{y})+I(\boldsymbol{x},\boldsymbol{y}|\B{H})-I(\boldsymbol{H},\boldsymbol{y}|\B{x})\\
&\geq I(\boldsymbol{x},\boldsymbol{y}|\B{H})-I(\boldsymbol{H},\boldsymbol{y}|\B{x})\\
&=\textrm{E}_{\boldsymbol{x},\boldsymbol{H}}\left[\sum_{\boldsymbol{y}}P(\boldsymbol{y}|\boldsymbol{x},\boldsymbol{H})\textrm{ln}\frac{\textrm{E}_{\boldsymbol{H}}[P(\boldsymbol{y}|\boldsymbol{x},\boldsymbol{H})]}{\textrm{E}_{\boldsymbol{x}}[P(\boldsymbol{y}|\boldsymbol{x},\boldsymbol{H})]}\right],
\label{transinfo_lower}
\end{aligned}
\end{equation}
where we can express each of the conditional probabilities $P(\boldsymbol{y}|\boldsymbol{x},\boldsymbol{H})$ as the product of the conditional probabilities on each receiver dimension, since all of the real and imaginary components of the receiver noise $\boldsymbol{\eta}$ are statically independent with power $\frac{1}{2}$
\begin{equation}
\begin{aligned}
P(\boldsymbol{y}|\boldsymbol{x},\boldsymbol{H})&=\prod_{c\in\{R,I\}}\prod_{i=1}^{N}P(y_{c,i}|\boldsymbol{x})\\
&=\prod_{c\in\{R,I\}}\prod_{i=1}^{N}\Phi\left(y_{c,i}[\boldsymbol{H}\boldsymbol{x}]_{c,i}\sqrt{2\rho}\right),
\end{aligned}
\end{equation}
with $\Phi(x)=\frac{1}{\sqrt{2\pi}}\int_{-\infty}^{x}e^{-\frac{t^2}{2}}dt$ is the cumulative normal distribution function. Evaluating the lower bound in (\ref{transinfo_lower}), even numerically, is very difficult, except for some simple cases, like SISO block faded channels \cite{mezghaniisit2008}. Thus, in the following,  we shall derive a quadratic approximation of the mutual information.
\section{Second-order Expansion of the Mutual Information}
\label{section:mutual2}
In this section, we will elaborate on the second-order expansion of the
input-output mutual information (\ref{transinfo}) of the considered system in Fig.~\ref{downlink_figure} as
the signal-to-noise ratio goes to zero. We state the main result and we prove it at the end of the section.  
\newtheorem {theorem1}{Theorem}
\begin {theorem1}
\label{dualitytheorem}
Consider the one-bit quantized MIMO system in Fig.~\ref{downlink_figure} under an input distribution $p(\boldsymbol{x})$  satisfying  $\textrm{E}_{\boldsymbol{x}} [\left\|\boldsymbol{x}\right\|_4^{4+\epsilon}]<\delta$ for some finite constants $\epsilon,\delta>0$. Then, to the second order, the mutual information (in nats) between the inputs and the quantized outputs is given by
\begin{equation}
\begin{aligned}	
I(\boldsymbol{x},\boldsymbol{y})=& \frac{1}{2} \left(\frac{2}{\pi} \rho \right)^2  {\rm tr} \left\{{\rm E} \left[({\rm nondiag}({\rm E}[\B{H}\B{x}\B{x}^{\rm H}\B{H}^{\rm H}|\B{x}])   )^2 \right] \right. \\
&\left. - ( {\rm nondiag}({\rm E}[\B{H}{\rm E}[\B{x}\B{x}^{\rm H}]\B{H}^{\rm H}])  )^2  \right\} 
+\underbrace{\Delta I(\boldsymbol{x},\boldsymbol{y})}_{o(\rho^2)},
\label{transinfo_2order}
\end{aligned}
\end{equation}
where $\textrm{nondiag}(\boldsymbol{A})$ is obtained from $\boldsymbol{A}$ by setting all its diagonal entries to zero, and $\left\|\boldsymbol{a} \right\|_4^4$ is the 4-norm of $\boldsymbol{a}$ taken to the power 4 defined as $\sum_{i,c}a_{i,c}^4$.

\end {theorem1}
\subsection{Comments on Theorem \ref{dualitytheorem}}
\begin{figure}[h]
\begin{center}
\psfrag{N}[c][c]{$N$}
{\epsfig {file=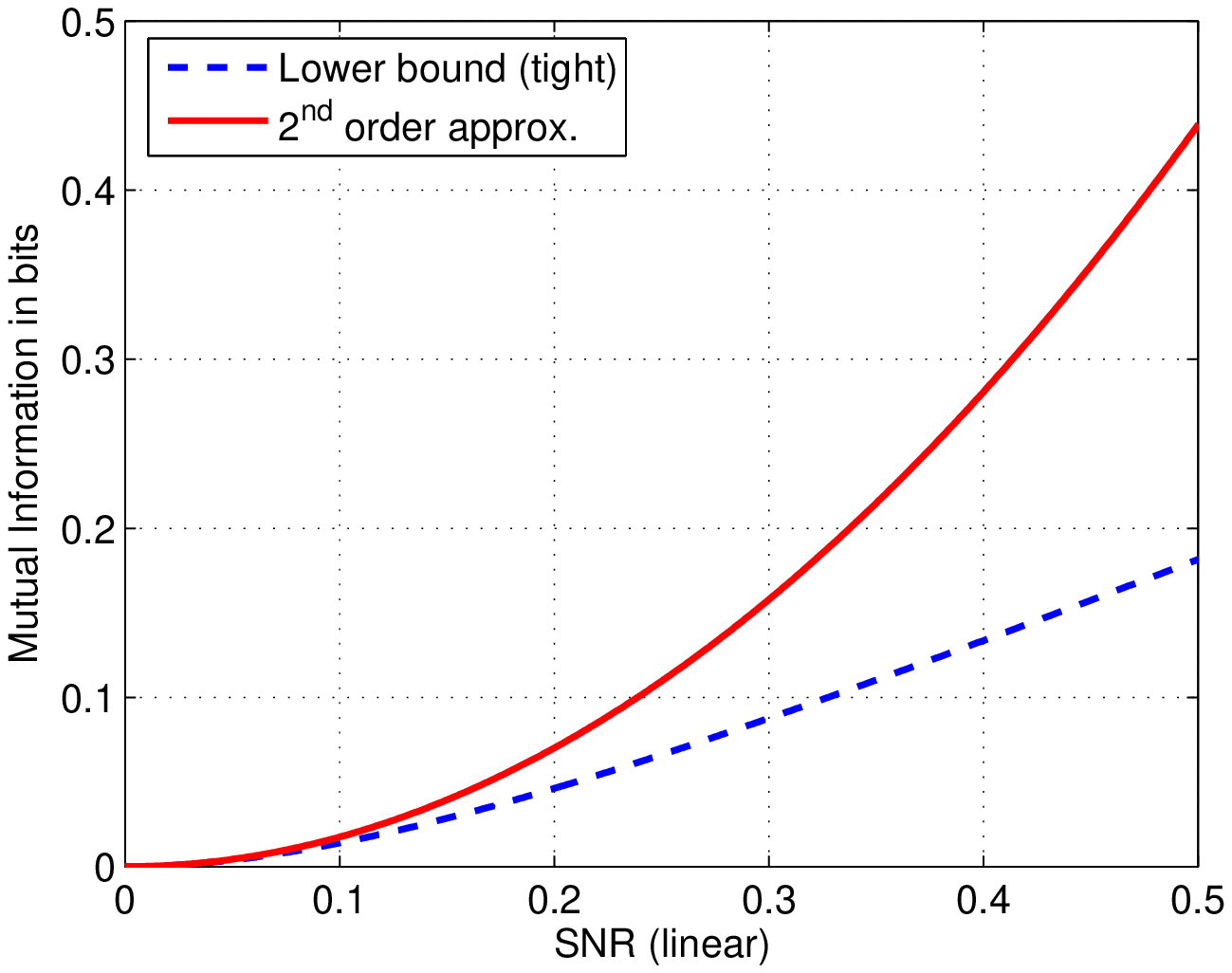, width = 7.8cm}}
\caption{Mutual information of the mono-bit block Rayleigh-faded SISO channel with block length 3.}
\label{mutual}
\end{center}
\end{figure}

As example, Fig.~\ref{mutual} illustrates the lower bound (\ref{transinfo_lower}) and the quadratic approximation (\ref{transinfo_2order}) computed for a  block faded SISO model with a coherence interval of 3 symbol periods ($\B{H}=h\cdot\mathbf{I}_3$, $h \sim \mathcal{CN}(0,1)$)  under QPSK signaling. Note that the lower bound (\ref{transinfo_lower}) is tight, since $I(\boldsymbol{H},\boldsymbol{y})=0$ in this case \cite{mezghaniisit2008}.
\vspace{-0.3cm}

\vspace{-0.2cm}
A similar result was derived by Prelov and Verd\`u in \cite{prelov} for the soft output $\B{r}$
\begin{equation}
\begin{aligned}	
I(\boldsymbol{x},\boldsymbol{r})=& \frac{1}{2} \rho^2  {\rm tr} \left\{{\rm E} \left[({\rm E}[\B{H}\B{x}\B{x}^{\rm H}\B{H}^{\rm H} |\B{x}]  )^2 \right] \right. \\
&\left. - ( {\rm E}[\B{H}{\rm E}[\B{x}\B{x}^{\rm H}]\B{H}^{\rm H}] )^2  \right\} 
+o(\rho^2),
\label{transinfo_2order_unq}
\end{aligned}
\end{equation}
 where we identify a power penalty of $\frac{\pi}{2}$ due to quantization and  that the diagonal elements of ${\rm E}_{\B{H}}[\B{H}\B{x}\B{x}^{\rm H}\B{H}^{\rm H}]$ do not contribute to the mutual information in the hard-decision system. This is because of the fact that no mutual information can be extracted from the amplitude of each two-level received signal. That means that the channel coefficients have to be correlated, otherwise ${\rm E}_{\B{H}}[\B{H}\B{x}\B{x}^{\rm H}\B{H}^{\rm H}]$ is diagonal and the mutual information is zero. Nevertheless in most systems of practical interest correlation between the channel coefficients, whether temporal or spatial, exists even in a multipath rich mobile environments, thus  ${\rm E}_{\B{H}}[\B{H}\B{x}\B{x}^{\rm H}\B{H}^{\rm H}]$ is rather a dense matrix, whose Frobenius norm is dominated by the off-diagonal elements rather than the diagonal entries. Thus the low SNR penalty due to hard-decision is nearly 1.96 dB for almost all practical channels. 
\subsection{Proof of Theorem \ref{dualitytheorem}}
As first step, we compute the conditional entropy $H(\B{y}|\B{x})$ up to the second order of the SNR using a quadratic Taylor expansion of $P(\B{y}|\B{x})$ around $\rho=0$
\begin{equation}
\begin{aligned}
P(\B{y}|\B{x})\approx P(\B{y}|\B{x})_{\rho=0}+\rho P'(\B{y}|\B{x})_{\rho=0} +\frac{\rho^2}{2}P''(\B{y}|\B{x})_{\rho=0},
\end{aligned}
\end{equation}
where $P'(\B{y}|\B{x})_{\rho=0}$  and $P''(\B{y}|\B{x})_{\rho=0}$ are the first and second derivatives of $P(\B{y}|\B{x})$ with respect to $\rho$. \\
Then, its logarithm can be approximated as\\
\begin{equation}
\begin{aligned}
\ln P(\B{y}|\B{x})\approx & \ln P(\B{y}|\B{x})_{\rho=0}+\rho  \frac{P'(\B{y}|\B{x})_{\rho=0}}{P(\B{y}|\B{x})_{\rho=0}} +\\
& \frac{\rho^2}{2} \left[ \frac {P''(\B{y}|\B{x})_{\rho=0}}{P(\B{y}|\B{x})_{\rho=0}} - \left(\frac {P'(\B{y}|\B{x})_{\rho=0}}{P(\B{y}|\B{x})_{\rho=0}} \right)^2\right].
\end{aligned}
\end{equation}
Therefore,
\begin{equation}
\begin{aligned}
&H(\B{y}|\B{x})\approx -\sum_{\B{y}} \Bigg[ P(\B{y}|\B{x})_{\rho=0} \ln P(\B{y}|\B{x})_{\rho=0} \\
&~~~~~~~~~~~~-\rho P'(\B{y}|\B{x})_{\rho=0}(1+\ln(P(\B{y}|\B{x})_{\rho=0}))\\
&-\frac{ \rho^2}{2} \left( \!\!P''(\B{y}|\B{x})_{\rho=0}(1+\ln(P(\B{y}|\B{x})_{\rho=0}))+ \frac {P'(\B{y}|\B{x})_{\rho=0}^2}{P(\B{y}|\B{x})_{\rho=0}} \right)\!\! \Bigg].
\end{aligned}
\end{equation}
Clearly $P(\B{y}|\B{x})_{\rho=0}=1/4^N$, and since $P(\B{y}|\B{x})$ is a probability mass function, i.e. $\sum_{\B{y}} P(\B{y}|\B{x})=1$, we have
\begin{equation}
\begin{aligned}
\sum_{\B{y}}P'(\B{y}|\B{x})=\sum_{\B{y}}P''(\B{y}|\B{x})=0.
\end{aligned}
\end{equation}
Hence, to the second order, the conditional entropy has the following quadratic approximation:
\begin{equation}
\begin{aligned}
H(\B{y}|\B{x}) \approx 2N \ln{2}-\frac{\rho^2}{2} \sum_{\B{y}} 4^N  {\rm E}_{\B{x}}\left[P'(\B{y}|\B{x})_{\rho=0}^2 \right]. 
 \end{aligned}
\end{equation}
In a similar way, we can show that
\begin{equation}
\begin{aligned}
H(\B{y}) \approx 2N \ln{2}-\frac{\rho^2}{2} \sum_{\B{y}} 4^N \left({\rm E}_{\B{x}}[ P'(\B{y}|\B{x})_{\rho=0}] \right)^2.
 \end{aligned}
\end{equation}
So, combining these results gives the quadratic expansion of the mutual information
\begin{equation}
\begin{aligned}
I(\B{x},\B{y})&=H(\B{y})-H(\B{y}|\B{x}) \\
& \approx \frac{\rho^2}{2}  4^N \!  \sum_{\B{y}} \! \left({\rm E}_{\B{x}}\!\left[ P'(\B{y}|\B{x})_{\rho=0}^2 \right] \!\!- \!\! \left({\rm E}_{\B{x}}[ P'(\B{y}|\B{x})_{\rho=0}] \right)^2 \right)\!\!.
 \end{aligned}
 \label{inter_approx}
\end{equation}
Now, the required first derivative is given by this result.
\newtheorem {lemma1}{Lemma}
\begin {lemma1}
The first derivative of the conditional probabilities in the hard-decision system with respect to $\rho$ is
\begin{equation}
\begin{aligned}
P'(\B{y}|\B{x})_{\rho=0}=\frac{1}{4^N}\frac{1}{\pi} \B{y}^{\rm H} \cdot {\rm nondiag}( {\rm E}[\B{H}\B{x}\B{x}^{\rm H}\B{H}^{\rm H}] ) \cdot \B{y}.
 \end{aligned}
\end{equation}
\label{lemma1}
\end {lemma1}
\begin{proof} 
Let us first expand the integral in (\ref{cond_pro}) to the first order. To this end we use the following approximation of (\ref{unq_cond_pro})\footnote{The following identities are useful: $\frac{\partial }{ \partial \rho} {\rm det} \B{Q}= {\rm det} \B{Q} \cdot {\rm tr}(\B{Q}^{-1} \frac{\partial \B{Q}}{ \partial \rho}) $ and $\frac{\partial  \B{Q}^{-1}}{ \partial \rho}=-  \B{Q}^{-1}\frac{\partial \B{Q}}{ \partial \rho} \B{Q}^{-1} $.}
\begin{equation}
\begin{aligned}
p(\boldsymbol{y} \bullet \B{r}|\B{x}) \approx \frac{\exp(-\| \B{r}\|^2)}{\pi^N} \left(1-\rho {\rm tr}({\rm E}_{\B{H}}[\B{H}\B{x}\B{x}^{\rm H} \B{H}^{\rm H}])\right. \\
 \left. + \rho {\rm tr} \{  (\boldsymbol{y} \bullet \B{r})(\boldsymbol{y} \bullet \B{r})^{\rm H}  {\rm E}_{\B{H}}[\B{H}\B{x}\B{x}^{\rm H} \B{H}^{\rm H}]  \}\right).
 \end{aligned}
 \label{unq_cond_prob_app}
\end{equation}
Afterward, it can be shown that
\begin{equation}
\begin{aligned}
\int_0^{\infty}\!\!\!\!\!\!\!\cdots\!\!\!\int_0^{\infty}  \frac{{\rm e}^{-\| \B{r}\|^2}}{\pi^N} (y_i \bullet r_i)(y_j \bullet r_j)^{\rm *} {\rm d}\B{r}=\left\{
\begin{array}{ll}
\frac{1}{4^N}	 & \textrm{for } i=j\\
	\frac{y_iy_j^*}{\pi4^N}	 & {\rm else.}
\end{array}
  \right. 
 \end{aligned}
\end{equation}
Thus,
\begin{equation}
\begin{aligned}
&\int_0^{\infty}\!\!\!\!\!\!\!\cdots\!\!\!\int_0^{\infty}  \frac{{\rm e}^{-\| \B{r}\|^2}}{\pi^N} {\rm tr}((\B{y} \bullet \B{r})(\B{y} \bullet \B{r})^{\rm H}   {\rm E}_{\B{H}}[\B{H}\B{x}\B{x}^{\rm H} \B{H}^{\rm H}]) {\rm d}\B{r}=\\
& \frac{1}{4^N}  {\rm tr} \{ ( \textbf{I}_N + \frac{1}{\pi}{\rm nondiag}(\B{y}\B{y}^{\rm H})){\rm E}_{\B{H}}[\B{H}\B{x}\B{x}^{\rm H} \B{H}^{\rm H}] \}= \\
&\frac{1}{4^N}  {\rm tr} \{{\rm E}\!_{\B{H}}\![\B{H}\!\B{x}\B{x}^{\rm H}\! \B{H}^{\rm H}] \}  \!+\! \frac{1}{4^N \! \pi} \B{y}^{\rm H}{\rm nondiag}({\rm E}\!_{\B{H}}\![\B{H}\!\B{x}\B{x}^{\rm H}\! \B{H}^{\rm H}])\B{y},
 \end{aligned}
 \label{int_approx}
\end{equation}
since ${\rm tr}({\rm nondiag}(\B{A})\B{B})= {\rm tr}(\B{A}{\rm nondiag}(\B{B}))$ for any two quadratic matrices $\B{A}$ and $\B{B}$. Integrating the quadratic approximation in (\ref{unq_cond_prob_app}) over the positive orthant by making use of (\ref{int_approx}) and a differentiation with respect to $\rho$  leads to the result  of the lemma.
\end{proof} 
Next, we plug the result of Lemma \ref{lemma1} into the required expression of the mutual information in (\ref{inter_approx}); and finally, we evaluate the summation over all possible quantized outputs $\B{y}$ according to the following identity 
\begin{equation}
\begin{aligned}
\frac{1}{4^N} \sum_{\B{y}\in \{\pm 1\pm {\rm j}\}^N}  (\B{y}^{\rm H}\B{P}\B{y})^2=4 {\rm tr} (\B{P}^2),
\end{aligned}
\end{equation}
which can verified for any  zero-diagonal matrix $\B{P}$ by expanding the left-hand side and identifying the nonzero terms. This yields  the result stated by the theorem and completes the proof. \\
Note that the condition $\textrm{E}_{\boldsymbol{x}} [\left\|\boldsymbol{x}\right\|_4^{4+\epsilon}]<\delta$ for some finite constants $\epsilon,\delta>0$ is necessary, so that the rest term of the expansion having the order
\begin{equation}
\Delta I(\boldsymbol{x},\boldsymbol{y})=\textrm{E}_{\boldsymbol{x}}[o(\left\|\boldsymbol{x}\right\|_4^4\rho^2)],
\end{equation}
satisfies
\begin{equation} 
 \lim_{\rho\rightarrow 0}\frac{\Delta I(\boldsymbol{x},\boldsymbol{y})}{\rho^2}=0,
 \end{equation} 
 since
\begin{eqnarray}
\begin{aligned}
\!\Delta I(\boldsymbol{x},\boldsymbol{y})&=	\textrm{E}_{\boldsymbol{x}}[o(\left\|\boldsymbol{x}\right\|_4^4\rho^2)] \nonumber  \\
	&\leq  \textrm{E}_{\boldsymbol{x}}[(\left\|\boldsymbol{x}\right\|_4^4\rho^2)^{1+\frac{\epsilon'}{4}}], \textrm{ for some } \epsilon' \in ]0,\epsilon] \nonumber  \\
	&\leq \textrm{E}_{\boldsymbol{x}}[\left\|\boldsymbol{x}\right\|_4^{4+\epsilon'}]\rho^{2+\frac{\epsilon'}{2}} \nonumber  \\
		&\leq \textrm{E}_{\boldsymbol{x}}[\left\|\boldsymbol{x}\right\|_4^{4+\epsilon}]^{\frac{4+\epsilon'}{4+\epsilon}}\rho^{2+\frac{\epsilon'}{2}}    \textrm{(H\"older's inequality)} \nonumber\\
			&\leq \delta^{\frac{4+\epsilon'}{4+\epsilon}}\rho^{2+\frac{\epsilon'}{2}} = o(\rho^2). \nonumber
	\end{aligned}
\end{eqnarray}

\section{Application to SIMO Channels with Delay Spread and Receive Correlation}
\label{section:receiver}
In this section, we use the result from Theorem \ref{dualitytheorem} to compute the low SNR mutual information of a frequency-selective single input multi-output (SIMO) channel with delay spread and receive  correlation. The unquantized output of the considered model at time $k$ is 
\begin{equation} 
\B{r}_k= \sqrt{\rho}\sum_{t=0}^{T-1} \B{h}_k[t] x_{k-t}  +\B{\eta}_k \in \mathbb{C}^{N}, 
\label{convhx}
\vspace{-0.1cm}
\end{equation}
where the noise process $\{\B{\eta}_k\}$ is i.i.d. in time and space, while the $T$ fading processes $\{\B{h}_k[t]\}$ in each tap $t$ are assumed to be independent zero-mean proper complex Gaussian processes. Furthermore, we assume a separable temporal spatial correlation model, i.e. 
\begin{equation}
{\rm E}[\B{h}_k[t] \B{h}_{k'}[t']^{\rm H}]= \B{R} \cdot r(k-k') \alpha_t \delta[t-t'].
\end{equation}
Here, $\B{R}$ denotes the receive correlation matrix, $r(k)$ is the autocorrelation function of the fading process, and the scalars $\alpha_i$ represents the delay power profile. These correlations parameters can be normalized so that
\vspace{-0.1cm} 
\begin{equation}
{\rm tr}(\B{R})=N,~ r(0)=1 \textrm{ and } \sum_{t=0}^{T-1} \alpha_t=1.
\label{normalization}
\vspace{-0.1cm} 
\end{equation} 
In other words, the energy
in  each receive antenna's impulse response equals one on
average. 
On the other hand we assume the the transmit signal $x_k$ is subjected to an average power constraint ${\rm E}[\| x_k\|^2]\leq 1$ and a peak power constraint $|x_k|^2 \leq \beta$, $\forall k$, with $\beta \geq 1$. It should be pointed out that a peak power constraint constitutes a stronger condition than necessary for the validity of Theorem~\ref{dualitytheorem}, involving just a fourth-order moment constraint on the input.  
We consider now a time interval of length $n$ (a block of $n$ transmissions) and introduce conveniently the following notation. Collect a vector sequence $\boldsymbol{y}_k$ of length $n$ into the vector $\boldsymbol{y}$ as
\begin{equation}
\boldsymbol{y}^{\!(\!n\!)}=\left[\boldsymbol{y}^\textrm{T}_{n-1},\ldots,\boldsymbol{y}^\textrm{T}_0 \right]^\textrm{T},
\end{equation}
and form the block cyclic-shifted matrix $\boldsymbol{H}^{\!(\!n\!)}\in \mathbb{C}^{(N n)\times n}$,
\begin{equation}
\boldsymbol{H}^{\!(\!n\!)}\!=\!\left[
\begin{array}{ccccc}
\!\!\boldsymbol{h}_{n-1}[0]\!\!\!& \cdots &\!\! \boldsymbol{h}_{n-1}[T\!-\!1]\!\!&0&\cdots\\
&\ddots& &&\ddots\\
\ddots	&&  &\ddots&\\
	\cdots &\!\!\boldsymbol{h}_{0}[T\!-\!1]\!\!\!&0~~~~~\cdots&0& \!\!\boldsymbol{h}_{0}[0] \!\!
\end{array}
\right].
\end{equation}
With 
\begin{equation}
\boldsymbol{x}^{\!(\!n\!)}=\left[x_{n-1},\ldots,x_{0}\right]^\textrm{T},
\end{equation}
and $\boldsymbol{\eta}^{\!(\!n\!)}$ and $\boldsymbol{r}^{\!(\!n\!)}$ defined similar to $\boldsymbol{y}^{\!(\!n\!)}$, the following unquantized space-time model may be formulated as a (loose) approximation of (\ref{convhx})
\vspace{-0.1cm} 
\begin{equation} \boldsymbol{r}^{\!(\!n\!)}=\boldsymbol{H}^{\!(\!n\!)}\boldsymbol{x}^{\!(\!n\!)}+\boldsymbol{\eta}^{\!(\!n\!)}.
\vspace{-0.1cm} 
\end{equation}
\subsection{Asymptotic Achievable Rate}
Now, we elaborate on the asymptotic information rate of this channel setting. First we establish an upper bound and than we examine its achievability.
\newtheorem {proposition1}{Proposition}
\begin {proposition1}
If $r(k)$ is square-summable, then the mutual information of the described space-time model admits, for any distribution fulfilling the average and peak power constraints, the limiting upper bound
\vspace{-0.1cm} 
\begin{equation} 
\begin{aligned}
\lim_{n \rightarrow \infty }\frac{1}{n}I(\boldsymbol{x}^{\!(\!n\!)},\boldsymbol{y}^{\!(\!n\!)}) \leq  \left(\frac{2}{\pi} \rho \right)^2  U(\beta),
 \end{aligned}
 \label{upper}
\end{equation}
where
\begin{equation} 
 \begin{aligned}
 U(\beta)=\left\{
\begin{array}{ll}
\beta\cdot \mu +(\beta-1)\sigma & \textrm{for } \beta(\mu+\sigma)\geq 2 \sigma \\
\beta^2\frac{(\mu+\sigma)^2}{4\sigma} & \textrm{else, } 
\end{array}
 \right.
 \end{aligned}
\end{equation}
\begin{equation} 
 \begin{aligned}
 \mu={\rm tr}(\B{R}^2)\sum_{k=1}^{\infty}r(k)^2
 \end{aligned}
 \label{lambda}
\end{equation}
and
\begin{equation} 
 \begin{aligned}
 \sigma=\frac{1}{2}{\rm tr}(({\rm nondiag}(\B{R}))^2).
 \end{aligned}
 \label{sigma}
\end{equation}
\label{proposition1}
\end {proposition1}
\begin{proof} 
Due to the peak power constraints, conditions of Theorem~\ref{dualitytheorem} are satisfied; thus the second order approximation (\ref{transinfo_2order}) is valid. A tight upper bound is obtained by looking at the maximal value that can be achieved by  the expression (\ref{transinfo_2order}) up to the second order. We do this  in two steps. We  first maximize the trace expression in (\ref{transinfo_2order}) under a prescribed average power per symbol $\gamma$. The maximum can be, in turn, upper bounded by the supremum of the first term minus the infimum of the second term under the prescribed average power and the original peak power constraint. After that, we perform an optimization over the parameter $\gamma$ itself. That is 
\begin{equation}
\begin{aligned}	
&I(\boldsymbol{x}^{\!(\!n\!)},\boldsymbol{y}^{\!(\!n\!)})\leq \frac{1}{2} \left(\frac{2}{\pi} \rho \right)^2  \max_{0\leq\gamma\leq 1}   \Bigg\{ \\
&\!\!\! \sup_{ \stackrel {|x_k|\leq \beta, \forall k}{{\rm E}[\|\B{x}^{\!(\!n\!)}\|^2]= \gamma n}
 } \!\!\!\! \!\!\!\!{\rm tr}\! \left(\!{\rm E}\! \left[\!({\rm nondiag}({\rm E}[\B{H}^{\!(\!n\!)}\B{x}^{\!(\!n\!)}\B{x}^{\!(\!n\!),{\rm H}}\B{H}^{\!(\!n\!),{\rm H}}|\B{x}])   )^2 \right] \right) \\
&\!\!\! -\!\!\!\!\!\!\!\! \inf_{ \stackrel {|x_k|\leq \beta, \forall k}{{\rm E}[\|\B{x}^{\!(\!n\!)}\|^2\!]= \gamma n}}\!\! \!\!\!\!\!\!{\rm tr} \!\left(\!( {\rm nondiag}({\rm E}[\B{H}^{\!(\!n\!)}{\rm E}\![\B{x}^{\!(\!n\!)}\B{x}^{\!(\!n\!),{\rm H}}]\B{H}^{\!(\!n\!),{\rm H}}])  )^2 \!  \right) \!\! \!\Bigg\}
\!\!+\!\!{o(\rho^2)}.
\label{transinfo_upper}
\end{aligned}
\end{equation}
It turns out that the supremum of first trace term is achieved when all $x_k$ inputs take,  simultaneously during the considered time interval, either the value zero, or the peak value $\beta$ with a duty cycle of $\gamma\beta^{-1}$. On the other hand, the infimum of the second trace expression is obtained, under the prescribed average power, when ${\rm E}[\B{x}\B{x}^{\rm H}]=\gamma {\bf I}_n$. Calculation shows that
\begin{equation}
\begin{aligned}	
\!\!\frac{1}{n}I(\!\boldsymbol{x}^{\!(\!n\!)},\boldsymbol{y}^{\!(\!n\!)}\!)\!\! \leq &\!\! \max_{0\leq  \gamma \leq 1} \!\! \left(\!\frac{2}{\pi} \rho \!\right)^{\!\!2}\!\!\left(\sum_{t=0}^{T-1}\! \alpha_t \!\!\right)^{\!\!\!2} \!\! \Bigg[\!\gamma\beta \!\cdot\!\underbrace{{\rm tr}(\!\B{R}^2)\!\!\sum_{k=1}^{n-1}(1\!-\!\frac{k}{n})r(k)^2}_{\mu^{\!(\!n\!)}}  \\
&\!\!+  \gamma(\beta-\gamma) \underbrace{\frac{1}{2}{\rm tr}(({\rm nondiag}(\B{R}))^2) r(0) }_{\sigma}\Bigg] +o(\rho^2).
\end{aligned}
\end{equation}
Now, the maximization over $\gamma$ delivers 
\begin{equation}
\begin{aligned}	
\gamma_{\rm opt}= \min \{1,\beta \frac{\mu^{\!(\!n\!)}+\sigma}{2\sigma}\},
\end{aligned}
\end{equation}
Thus, taking the limit $n \rightarrow \infty$ yields $\mu^{\!(\!n\!)} \rightarrow \mu$ as in  (\ref{lambda}) and by the normalizations in (\ref{normalization}), we end up by the result of the proposition.\footnote{Observe that $\mu$ and $\sigma$ are indicators for the temporal and spatial coherence, respectively.}
\end{proof}
We next turn to the question, whether the upper bound  suggested by Proposition \ref{proposition1} is achievable. A closer examination of (\ref{transinfo_upper}) demonstrates that the upper bound could be achieved if the input distribution would satisfy, for any time instants $k$ and $l$ within the block of length $n$
\begin{equation}
\begin{aligned}	
x_k x_l^*= a {\rm e}^{{\rm j}\Omega(k-l)},
\end{aligned}
\end{equation}
with some $\Omega\neq 0$ and $a\in\{0,\beta \}$, while having ${\rm E}[\B{x}^{\!(\!n\!)}\B{x}^{\!(\!n\!),{\rm H}}]=\gamma {\bf I}_{n}$ as already mentioned in the proof of Proposition \ref{proposition1}. Clearly an \emph{on-off frequency-shift keying} (OOFSK) modulation for the input, as follows,  can fulfill these conditions  
\begin{equation}
\begin{aligned}
x_k=Z\cdot {\rm e}^{{\rm j}k\Omega},~~k\in\{1,\ldots,n\},	 
\end{aligned}
\end{equation}
where $Z$ takes the value $\sqrt{\beta}$ with probability $\gamma_{\rm opt} \beta^{-1}$ and zero with probability $(1-\gamma_{\rm opt} \beta^{-1})$, and $\Omega$ is uniformly distributed over the set $\{\frac{2\pi}{n} ,\ldots,\frac{2\pi(n-1)}{n} \}$. This is exactly similar to the results  of  the unquantized case \cite{sethuraman,telatar}. 
\subsection{Discussions}
\begin{figure}[h]
\begin{center}
\psfrag{T=1}[c][c]{\footnotesize $T=1$}
\psfrag{T=5}[c][c]{\footnotesize $T=5$}
\psfrag{sigma/mu}[c][c]{ $\frac{\sigma}{\mu}$}
{\epsfig {file=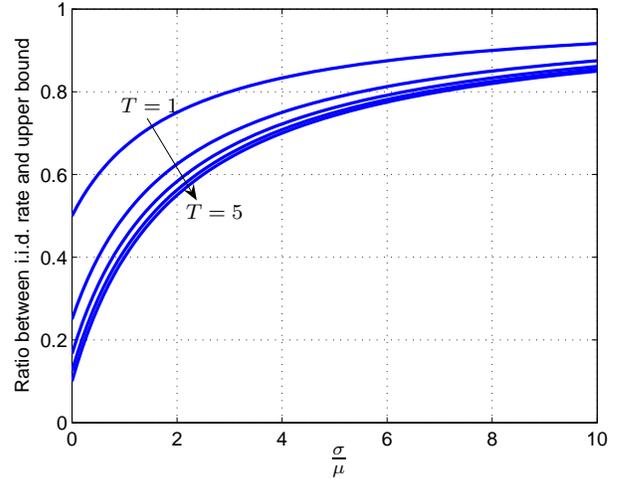, width = 9cm}}
\caption{Ratio of i.i.d. rate (\ref{iidrate}) and upper bound (\ref{upper}) vs. the spatial-to-temporal coherence ratio $\frac{\sigma}{\mu}$ for $\beta=2$ and uniform delay spread, i.e.  $\alpha_t=\frac{1}{T}$.}
\label{iid_rate_figure}
\end{center}
\vspace{-0.5cm}
\end{figure}

We observe from Proposition \ref{proposition1} that spreading the power into different taps does not affect the low SNR rate, while receiver correlation is beneficial due to two effects. First the mutual information increases with $\sigma$  defined in (\ref{sigma}) which is related to the norm of the off-diagonal elements of $\B{R}$. Second, under the normalization ${\rm tr}(\B{R})=N$, the Frobenius norm ${\rm tr}(\B{R}^2)$ increases with more correlation among the receive antenna, and consequently, higher rates at low SNR can be achieved due to relation (\ref{lambda}).
Besides we note that the achievability of the upper bound stated in Proposition \ref{proposition1}, as discussed previously, is obtained at the cost of burstiness  in frequency which may not agree with some 
specifications imposed on many systems.\footnote{Note that such observations holds also in the unquantized case \cite{sethuraman}.} Therefore, it is interesting to look at asymptotic rate of i.i.d. input symbols drawn from the set
$\{ -\sqrt{\beta},0, \sqrt{\beta} \}$. In that case it turns out by (\ref{transinfo_2order}) that
  \begin{equation}
\begin{aligned}
\frac{1}{n}I_{\rm IID}(\boldsymbol{x}^{\!(\!n\!)},\boldsymbol{y}^{\!(\!n\!)}) \!\!\approx \!\!& \max_{0\leq  \gamma \leq 1} \left(\!\frac{2}{\pi} \rho \!\right)^{\!\!2}\!\! \left[ \gamma^2\mu^{\!(\!n\!)} \sum_{t=0}^{T-1} \alpha_t^2+  \gamma(\beta-\gamma) \sigma \right]\!\!.
\end{aligned}
\label{iidrate}
\end{equation}
Here, we observe that, contrary to to the FSK-like scheme, the mutual information with i.i.d input is negatively affected by the delay spread since $\sum_{t=0}^{T-1} \alpha_t^2\leq 1$ by the normalization (\ref{normalization}). Nevertheless for the case $\sigma \gg \mu$, i.e. low temporal correlation (high Doppler spread), the gap to the upper bound in Proposition \ref{proposition1}  vanishes, as demonstrated in Fig.~\ref{iid_rate_figure}.
\section{Conclusion}
\label{section:conclusion}
We derived an expression for the second-order expansion of the
mutual information for general MIMO channels with one-bit ADC for low SNR and general input distribution. Based  on this, we showed that the power penalty due to the 1-bit quantization is approximately equal $\frac{\pi}{2}$ (1.96 dB) at low SNR in practical channel settings. This confirms that low-resolution (or maybe 1-bit) sampling in the low SNR regime (or equivalently in the wideband limit) performs adequately, while reducing power consumption. Besides, it turns out that both spatial and temporal correlations are extremely beneficial at low SNR (even more than in the unquantized case). We studied the special case of multiple receive antennas with delay spread and correlation both in time and space in detail and obtained the asymptotic achievable rate  under average and peak power constraint.
\bibliographystyle{IEEEbib}
\bibliography{IEEEabrv,references}
\end{document}